\documentclass[twocolumn,showpacs, showkeys,preprinumbers,amsmath,amssymb, aps, prb]{revtex4}
\usepackage{graphicx}
\usepackage{dcolumn}
\usepackage{bm}
\usepackage[dvips]{color}

\begin{document}

\preprint{AIP/123-QED}

\title{Antiferromagnetic ordering in EuPtGe$_3$ single crystal}
\author{Neeraj Kumar}
\email{goyal@tifr.res.in}
\author{Pranab Kumar Das}
\author{Ruta Kulkarni}
\author{A. Thamizhavel}
\author{S. K. Dhar}
\affiliation{Department of Condensed Matter Physics and Material
Sciences, Tata Institute of Fundamental Research, Homi Bhabha Road,
Colaba, Mumbai 400 005, India}
\author{P. Bonville}
\affiliation{CEA Saclay, DSM/IRAMIS, Service de Physique de l'Etat Condens\'e, 91191 Gif-Sur-Yvette, France}

\date{\today}

\begin{abstract}

The magnetic properties of single crystalline  EuPtGe$_3$, crystallizing in the non-centrosymmetric BaNiSn$_3$-type crystal structure, have been studied by means of magnetisation, electrical resistivity, heat capacity and $^{151}$Eu M\"ossbauer spectroscopy. The susceptibility and heat capacity data indicate a magnetic transition at $T_{\rm N}$ = 11\,K. The M\"ossbauer data confirm this conclusion, but evidence a slight first order character of the transition. Analysing the magnetisation data using a mean field model with two antiferromagnetically coupled sublattices allows to explain some aspects of the magnetic behaviour, and to derive the first and second neighbour exchange integrals in EuPtGe$_3$.
\end{abstract}



\maketitle

\section{Introduction}

After the discovery of heavy fermion superconductivity and magnetic ordering in non-centrosymmetric CePt$_3$Si~\cite{bauer}, compounds with non-centrosymmetric structure have received a great deal of attention.  Recently, we reported the magnetic properties of EuPtSi$_3$ single crystal~\cite{Neeraj} which lacks inversion symmetry.  The magnetization data reveal that EuPtSi$_3$ undergoes two closely spaced successive antiferromagnetic transitions at 16 and 17\,K, respectively. The heat capacity and $^{151}$Eu M\"ossbauer measurements confirmed that EuPtSi$_3$ undergoes a cascade of transitions from the paramagnetic phase to an incommensurate amplitude modulated followed by a commensurate equal moment magnetic structure. This type of cascading transition in Eu systems is not uncommon as it has been found in layered semimetallic EuAs$_3$ \cite{Chattopadhyay} between $T_{\rm N1}$=11.3\,K and $T_{\rm N2}$=10.26\,K and in EuPdSb \cite{Bonville} between $T_{\rm N1}$=18\,K and $T_{\rm N2}$=12\,K. Since Ge lies in the same group as Si, it was interesting to explore how the  replacement of Si by Ge in EuPtSi$_3$ affects the magnetic properties. In the present work, we report the synthesis and the magnetic properties of single crystalline EuPtGe$_3$, which also crystallizes in the non-centrosymmetric BaNiSn$_3$ type crystal structure, like its Si analogue. We also show that mean field theory allows to account for some of the magnetic properties of EuPtGe$_3$.

\section{Experiment}

Single crystals of EuPtGe$_3$ were grown by the high temperature solution growth method using Sn flux. The growth procedure is exactly the same as described in Ref.\onlinecite{Neeraj}. The typical dimensions of the obtained crystals were 2~$\times$3~$\times$~3\,mm$^3$. The single crystal was found to be stable in air. A Huber-Laue diffractometer was used to orient the single crystal along the crystallographic directions, namely the [100],[110] and [001] axes. Powder x-ray diffraction was performed in a Phillips Pan-Analytical x-ray diffractometer which uses Cu-$K_{\alpha}$ radiation. The actual crystal composition was confirmed using a CAMECA SX100 Electron probe micro-analyser (EPMA). The dc magnetic susceptibility and the in-field magnetization measurements were performed in the temperature range 1.8\--300\,K using a Quantum Design superconducting quantum interference device (SQUID) magnetometer and Oxford vibrating sample magnetometer (VSM). The temperature dependence of electrical resistivity in the range 1.8\--300\,K was measured using a home made dc electrical resistivity set up. The heat capacity was performed using a Quantum Design physical properties measurement system (PPMS). For these measurements, the single crystal was cut along the required directions by a spark erosion cutting machine. The M\"ossbauer measurements on the $^{151}$Eu isotope were made on a powder sample. They were performed using a commercial Sm$^*$F$_3$ $\gamma$-ray source, with a constant acceleration electromagnetic drive, in a standard He cryostat.

\section{Results and Discussion}

\subsection{Crystal Structure}
In order to study the anisotropic physical properties, a single crystal was oriented by Laue back reflection. Well defined Laue diffraction spots with four fold symmetry were obtained as expected for the tetragonal symmetry of the crystal lattice. Powder x-ray diffraction pattern revealed that the sample is almost single phase; a few extra weak lines were also seen which are due to the trace amounts of Sn (flux) present on the surface of the single crystal. Lattice parameters obtained from Rietveld analysis using the FullProf software package \cite{Carjaval} are $a$ = 4.4754\,\AA\ and $c$ = 10.1154\,\AA. These values are slightly higher than  those reported in Ref.\onlinecite{demchyna} ($a$=4.4633\,\AA and $c$=10.0625\,\AA) where a single crystal was isolated from a polycrystalline ingot. Compared to the corresponding values of $a$=4.2660\,\AA\ and $c$=9.8768\,\AA\ in EuPtSi$_3$, there is an expansion in both the $a$ and $c$-axis. This shows that the replacement of Si by Ge results in the expansion of the unit cell volume, due to the larger atomic size of Ge. 

\subsection{Magnetic Susceptibility}
\begin{figure}
\includegraphics[width=0.44\textwidth]{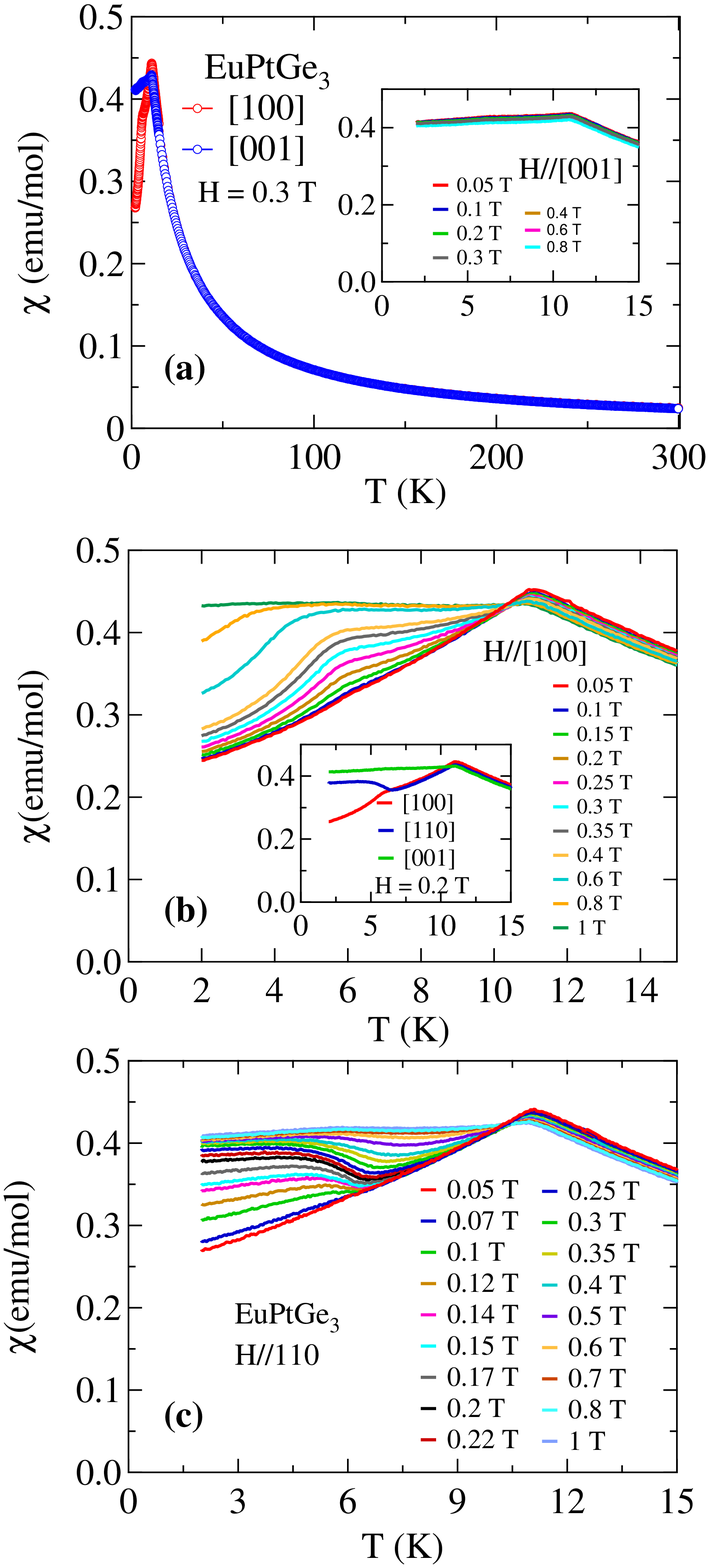}
\caption{\label{chiT}(Color online) (a) Main panel shows temperature dependence of the magnetic susceptibility measured in an applied field of 0.3\,T along the two principal directions in the temperature range from 1.8\--300~K. The inset shows the susceptibility for $H~\parallel$~[001] performed at various fields. (b) Main panel shows the susceptibility for $H~\parallel$~[100] performed at various fields and inset shows susceptibility at 0.2 T for $H~\parallel$~[100] and $H~\parallel$~[110]. (c) Magnetic susceptibility for $H~\parallel$~[110] performed at various fields.}
\end{figure}

The temperature dependence of the magnetic susceptibility in a field of 0.3\,T applied along [100] and [001] is shown in Fig.\ref{chiT} upper panel between 1.8 and 300\,K. An anomaly at $T_{\rm N} \simeq$11\,K reveals a transition to an antiferromagnetic (AF) phase, which is confirmed by the M\"ossbauer and heat capacity data (see below). The decrease in the AF ordering temperature with respect to EuPtSi$_3$, where $T_{\rm N} \simeq 17$\,K, may be attributed to the increase in Eu-Eu bond distance, which is equal to lattice constant $a$ caused by the larger size of Ge. Above 15\,K, magnetic susceptibility is isotropic and follows a Curie-Weiss(CW) law the along two principal axes. By fitting CW to the data between 20\--300\,K, $\mu_{\rm eff}$=7.77 and 8.11\,$\mu_B$, and $\theta_p$ =$-$5.4 and $-$7.6\,K for [100] and [001] axis, are obtained respectively. We obtain a polycrystalline average $\mu_{\rm eff}= 7.90 \mu_B$, which is close to calculated value of 7.94 $\mu_B$ characteristic of Eu$^{2+}$. The negative value of $\theta_p$'s points to a dominant AF first-neighbour Eu-Eu exchange; its absolute value somehow smaller than $T_{\rm N}$ suggests the presence of a small ferromagnetic second-neighbour exchange integral. By comparison, the situation is more complex in EuPtSi$_3$, where $\theta_p$ is positive along both [100] and [001] axes, which was attributed to a dominant ferromagnetic second neighbour exchange \cite{Neeraj}.

Whereas high temperature susceptibility is nearly isotropic significant anisotropy is observed at low temperature between the data along different axes (see Fig.\ref{chiT}). For very low field upto 0.05\,T $\chi_{[100]}$ and $\chi_{[110]}$ decrease from 11\,K down to 1.8\,K. On increasing the field an anomaly is seen in $\chi_{[100]}$ and $\chi_{[110]}$ which is centred around 6\,K. Below 6\,K, $\chi_{[100]}$ decreases more rapidly while as $\chi_{[110]}$ increases. At further higher fields anomaly at 6\,K vanishes and $\chi_{[100]}$ and $\chi_{[110]}$ become nearly temperature independent at 1\,T. Compared to these two axes field variation doesn't have much effect on $\chi_{[001]}$ which shows nearly temperature independent behaviour at all fields.

\begin{figure}[!]
\includegraphics[width=0.40\textwidth]{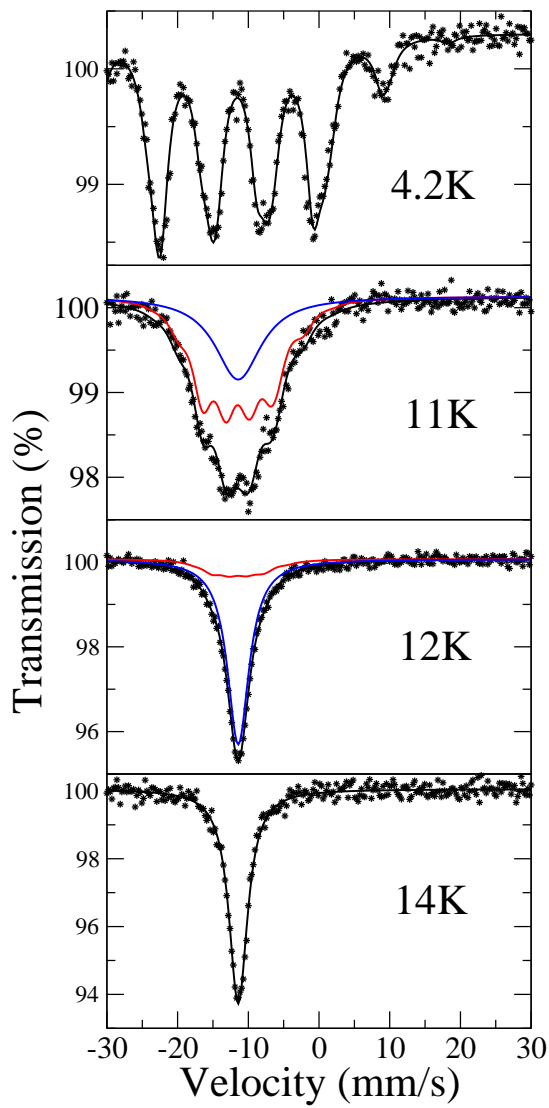}
\vspace{10pt}
\caption{\label{spmos}(Color online) $^{151}$Eu M\"ossbauer spectra in EuPtGe$_3$. Between 10.5\,K and 12\,K, a superposition of a hyperfine field spectrum (in red) and of a single line (in blue) is observed. The fits of the magnetically split spectra are to a single hyperfine field lineshape.}
\end{figure}

\subsection{M\"ossbauer Spectroscopy}

\begin{figure}[!]
\includegraphics[width=0.45\textwidth]{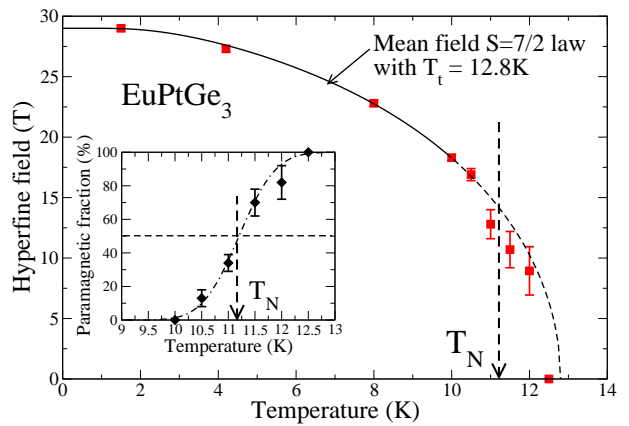}
\caption{\label{hhf}(Color online) Thermal variation of the hyperfine field in EuPtGe$_3$. The line is a mean-field calculation using the Brillouin function with S=7/2 and a transition temperature $T_t$=12.8\,K. Inset: Thermal variation of the paramagnetic fraction inferred from the spectra. The dashed line is a guide to the eye built from an $erf$ function.}
\end{figure}

In order to get further insight into the magnetic transition, we performed M\"ossbauer spectra on the isotope $^{151}$Eu in the temperature range 1.5-14\,K (see Fig.\ref{spmos}). From the base temperature up to 10\,K, a single magnetic hyperfine field spectrum is observed, characteristic of an equal moment magnetic structure. At 1.5\,K, the saturated hyperfine field is 29.0(2)\,T and the Isomer Shift is $-$11\,mm/s, typical for divalent Eu. As temperature is raised above 10\,K, a single line subspectrum (in blue in Fig.\ref{spmos}) grows at the expense of the magnetic subspectrum (in red in Fig.\ref{spmos}). This single line is characteristic of the paramagnetic phase; at 12.5\,K, the whole sample is paramagnetic. Such a coexistence of magnetic and paramagnetic domains in a narrow interval around $T_{\rm N}$ is observed when the transition has a certain degree of first order \cite{Bonville}. The temperature dependence of the hyperfine field is shown in Fig.\ref{hhf} and that of the fraction of the paramagnetic phase in the inset of Fig.\ref{hhf}. 

The hyperfine field decreases as temperature increases and, up to 10\,K, follows precisely a S=$\frac{7}{2}$ mean field law with a transition temperature $T_t$=12.8\,K. Above 10\,K, it somehow departs from the mean field law as the paramagnetic fraction grows. The ``mid-point'', where 50\% of the sample is paramagnetic, is reached very close to 11\,K, which is the actual N\'eel temperature, in very good agreement with the susceptibility measurements described above and with the specific heat data, to be described below. Therefore, the paramagnetic-to-AF transition in EuPtGe$_3$ possesses a slight degree of first order: the transition temperature is $T_{\rm N}\simeq$11\,K, whereas the mean field transition would occur at $T_t$=12.8\,K. 

\subsection{Heat Capacity}
\begin{figure}[h]
\includegraphics[width=0.45\textwidth]{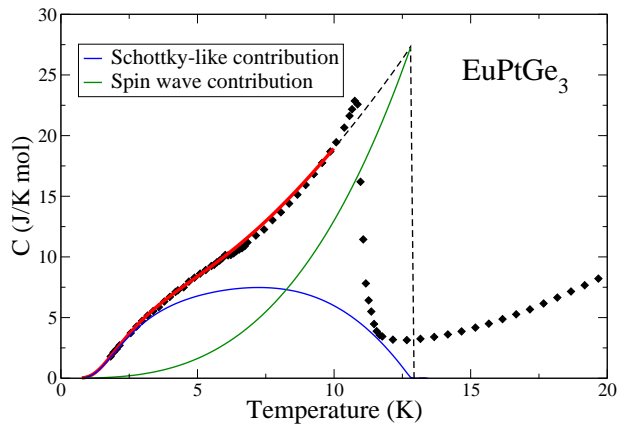}
\caption{\label{hc}(Color online) Temperature dependence of the heat capacity of EuPtGe$_3$. The red line is the sum of two contributions: a Schottky-like anomaly (in blue) and a magnon contribution varying as $T^3$ (in green). The calculations are performed within a mean field theory with transition temperature $T_t$=12.8\,K (see text), which holds up to 10\,K in EuPtGe$_3$, whence the truncation of the red line at 10\,K.}
\end{figure}

Figure \ref{hc} shows the heat capacity $C$ of EuPtGe$_3$ in zero field from 1.8\,K to 20\,K. The data show a sharp peak at 11\,K confirming the bulk nature of the magnetic transition. The $4f$ contribution to the heat capacity $C_{4f}$ could not be estimated as the non-magnetic reference compounds LaPtGe$_3$ or LuPtGe$_3$ are not available. However, up to 10-15\,K, the lattice contribution (phonons and conduction electrons) is expected to be relatively small, and one expects that $C_{4f}$ is well represented by the measured specific heat. For example in the non-magnetic iso-structural LaPtSi$_3$~\cite{Neeraj} $\gamma$=0.0043 \,J/mol K$^2$ and $\beta$=0.0002 \,J/mol K$^4$. The jump at $T_N$ is estimated to be 19.7\,J/mol K, which is close to the theoretical value (20.2 J/mol K) for a second order transition to an equal moment S=$\frac{7}{2}$ magnetic structure. In EuPtSi$_3$, by contrast, the jump is only 14.6\,J/mol K, which is closer to the value inferred for an amplitude modulated magnetic structure. The two main contributions to the heat capacity in a magnetically ordered phase are i) the Schottky-like anomaly $C_{\frac{7}{2}}$ due to the population of the 8 exchange split levels and ii) the magnon specific heat $C_{\rm sw}$ which is expected to follow a $T^3$ variation in a 3-dimensional antiferromagnet. Figure \ref{hc} shows a fit of the heat capacity to the sum of $C_{7/2}(T)$, $C_{\rm sw} = \alpha T^3$ and of C$_{el}$=$\gamma$T with $\alpha$=0.013\,J/mol K$^4$ and $\gamma$= 0.01\,J/mol K$^2$.  The agreement with experiment is good and reproduces the inflexion point observed near 6\,K. 

\subsection{Magnetization}
\begin{figure}[h]
\includegraphics[width=0.45\textwidth]{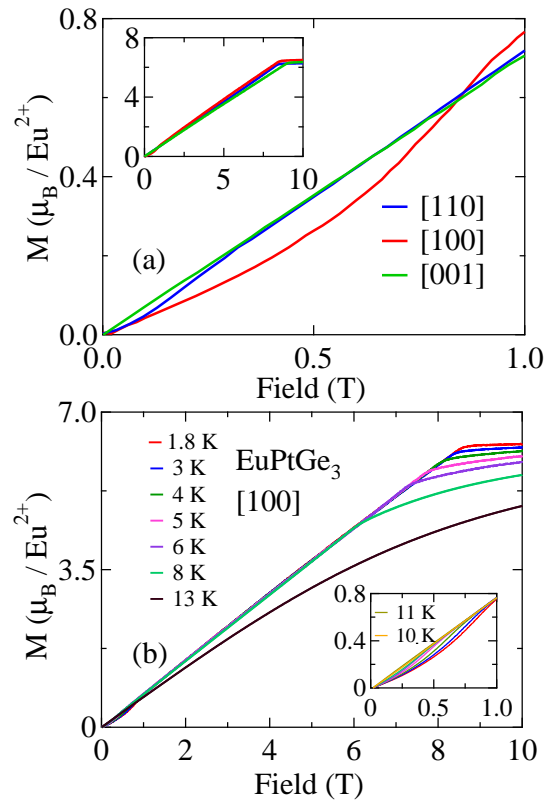}
\caption{\label{mH}(Color online) (a) Inset  shows field dependence of magnetization in the range 0\--10\,T. at 1.85\,K, for fields along [100],[110] and [001]. Main panel shows low field zoom showing the departure from linearity observed along [100] and [110]. (b) Isothermal Magnetization at various temperatures for a field applied along [100]. Inset: low field zoom showing the progressive disappearance of the non-linear behaviour with increasing temperature. Legend in the inset are same as main plot in addition to shown in inset.}
\end{figure}

In order to obtain further details about the magnetic structure, we performed magnetization measurements at 1.85\,K with a field applied along the [100], [110] and [001] axis (see Fig.\ref{mH},inset top panel). The magnetization is almost linear at low fields and saturates eventually at 8.5\,T along [100] and [110] and at 9\,T along [001]. This type of linear behaviour, and the saturation to a field-induced ferromagnetic phase above a spin-flip field which is almost independent from the direction, are characteristic of an AF structure with zero or very little anisotropy. The corresponding values of the spin-flip fields in the silicide are respectively 9.2 and 5.9\,T for $H~\parallel$~[100] and $H~\parallel$~[001] respectively. Thus, the magnetization data indicate a much lesser anisotropy in the germanide. The saturated moment($\mu_s$) is 6.5\,$\mu_B$/Eu$^{2+}$, which is 7\% lower than the theoretical $T=0$ moment of 7\,$\mu_B$. Similar value of $\mu_s$ was also seen in EuPtSi$_3$~\cite{Neeraj} where it was tentatively attributed to slight traces of Sn sticking to crystal surface. We speculate that a conduction electron polarization aligned antiparallel to Eu$^{2+}$ ions in these two compounds may also lead to a reduced $\mu_s$. A slight departure from a linear variation is observed below 1\,T for $H~\parallel$~[100] and below 0.3 T for $H~\parallel$~[110]. With increase in temperature, the departure from linearity weakens and it eventually disappears at 11\,K (see Fig.\ref{mH}(b) inset) for $H~\parallel$~[100]. Let us focus back our attention on the susceptibility data already shown in Fig.~\ref{chiT}. For $H~\parallel$~[001], the susceptibility is practically temperature independent which, together with the perfectly linear variation of the magnetization for this field direction, indicates that the easy magnetic axis of the AF structure is perpendicular to [001]. The susceptibility along [001] is thus of transverse type, and: $\chi_\perp \simeq$0.43\,emu/mol. At 0.05 T for fields applied along [100] or [110], the susceptibility increases smoothly from a value close to $\chi_\perp/2$ at 2\,K to $\chi_\perp$ at $T_{\rm N}$.  

We apply a mean field model with S=$\frac{7}{2}$ to interpret the magnetic data in EuPtGe$_3$. This model \cite{herpin} describes two AF sublattices, coupled through a first neighbour molecular field constant $\lambda_1$ ($<$0) and a second neighbour constant $\lambda_2$. The usual exchange integral $\cal J$ is linked to the molecular field constant by: ${\cal J}_i = (g \mu_B)^2 \lambda_i$, where $g=2$ for Eu$^{2+}$. The transverse susceptibility $\chi_\perp$ and the spin-flip field $H_{\rm sf}$ depend only on the $\lambda_1$ value: $\chi_\perp = \frac{1}{2 \vert \lambda_1 \vert}$ and $H_{\rm sf} = \frac{m_0}{\chi_\perp}$, where $m_0$=7\,$\mu_B$ is the saturated Eu$^{2+}$ moment. Taking $\lambda_1 = -0.64$\,T/$\mu_B$ reproduces well the measured values $H_{\rm sf} = 9$\,T (along [001]) and $\chi_\perp=0.43$\,emu/mol. Within this mean field model, the expressions for the transition temperature $T_t$ and for $\theta_{\rm p}$ are:

\begin{eqnarray}
 k_{\rm B} T_t & =  \vert \lambda_1-\lambda_2 \vert \ \frac{S+1}{3S} \ m_0^2 \cr\cr
 k_{\rm B} \theta_{\rm p} & =  (\lambda_1+\lambda_2) \ \frac{S+1}{3S} \ m_0^2.
\label{tn}
\end{eqnarray}

Taking $\lambda_2 = 0.25$\,T/$\mu_B$, i.e. of ferromagnetic type, yields $T_t$=12.8\,K, as inferred from the M\"ossbauer data ($T_t$ is slightly higher than the actual N\'eel temperature $T_{\rm N} \simeq 11$\,K due to the first order character of the transition). It also yields $\theta_p = -5.5$\,K, which agrees well with the value measured along [100], but which is smaller (in absolute value) than along [001] ($-$8.8\,K). We also include in the model a small crystalline anisotropy energy term ${\cal H}_{\rm anis} = D\ S_u^2$, with $D<0$, where $S_u$ is the component of the Eu$^{2+}$ spin along the easy axis determined by the unit vector {\bf u}. According to what was said above, we take the easy axis to lie in the (001) plane. Measuring the low temperature susceptibility (in the AF phase) for different orientations of the field in the (001) plane should allow the easy axis to be determined, since $\chi (T=0) = \chi_\perp \sin^2 \phi$, where $\phi$ is the angle between the field direction and the easy axis. For a crystal with a single magnetic domain, the susceptibility should vanish when the field is along the easy axis. However, the susceptibility values measured at 2\,K along [100] and [110] are the same and worth $\chi_\perp/2$, which is not possible for a single domain. So next we assume the presence of two AF domains, with perpendicular axes in the (001) plane, with angles $\phi$ and $\phi+ \frac{\pi}{2}$. The susceptibility is then: $\chi (T=0)= \frac{1}{2}\chi_\perp [\sin^2 \phi + \sin^2(\phi+\frac{\pi}{2})] = \frac{1}{2}\chi_\perp$. This is independent of $\phi$, i.e. of the field orientation, as observed experimentally. 

\begin{figure}[h]
\includegraphics[width=0.45\textwidth]{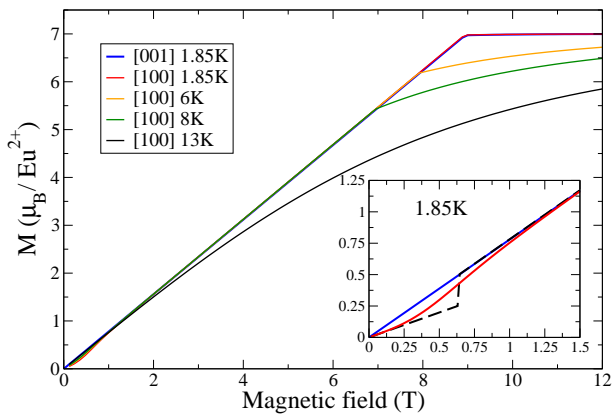}
\caption{\label{mHcalc}(Color online) Calculated field and temperature dependence of the magnetization in the range 0-12\,T, for a field applied along [001] and [100]. The parameters for the calculation are: $\lambda_1=-0.64$\,T/$\mu_B$, $\lambda_2=0.25$\,T/$\mu_B$ for the exchange, and $D=-0.01$\,K and $\phi=45^\circ$ for the anisotropy, i.e. an easy axis along [110]. Inset: low field zoom with same legend and parameters as the main figure, except the black dashed line which represents the calculation for $H~\parallel$~[100] with $\phi=0$, i.e. an easy axis along [100], and which evidences a spin-flop field $H_{flop} \simeq 0.68$\,T.}
\end{figure}

Thus, in the presence of twinning, the low field slope of the magnetization for $H~\parallel$~[100] and [110] should be $\chi_\perp/2$, as also observed experimentally. However, our model, with the presence of two domains, predicts  a first order spin-flip transition when the field is parallel to either [100] or [110] (with $\phi=0$), with a sharp jump of the magnetization at a critical field:

\begin{equation}
H_{flop} = \frac{m_0}{\mu_B} \ \sqrt{\vert \lambda_1 D \vert}.
\label{hflo}
\end{equation}
The magnetization curve should present a jump at $H_{flop} \simeq 0.68$\,T in agreement with expression (2) for D = -0.01 K (black dashed line in the inset of Fig. \ref{mHcalc}). However, no jump is observed experimentally. On the other hand, the low field magnetization ($<$ 1 T) at 1.8\,K along [100] and [110] is fairly well reproduced theoretically for two different values of D, for a wide range of angles of inclination between the field and the assumed easy axis direction. For example, in the inset of Fig.6, red curve is computed for $H~\parallel$~[100] with easy axis along [110], which matches with experimental data quite closely. As the temperature is increased the susceptibility shows an anomalous behaviour around 6 K in the field 0.05$<$H$<$1 T (Fig.~\ref{chiT}), which we are not able to reproduce using our model. This indicates that our simple model does not capture the full details of the magnetization of EuPtGe$_3$. More generally, Fig. \ref{mHcalc} shows the good overall agreement of our model with the magnetization data at different temperatures at relatively higher fields (for calculating these curves, we have assumed a saturation moment of 7$\mu_B$). 

\subsection{Electrical Resistivity}
\begin{figure}[h]
\includegraphics[width=0.45\textwidth]{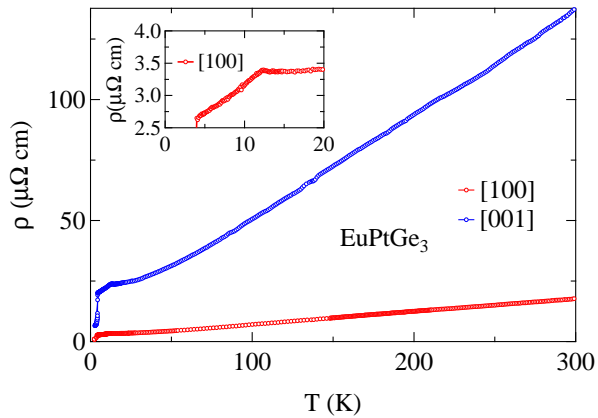}
\caption{\label{res}(Color online) Temperature dependence of electrical resistivity for $J \parallel $ [100] and [001] in the temperature range from 1.8 to 300\,K. The inset shows the low temperature part with $J\parallel$[100] .}
\end{figure}

The temperature dependence of the electrical resistivity is shown in Fig.~\ref{res} for a current parallel to the [100] and [001] directions. Magnitude of resistivity is typical for an intermetallic compound and decreases with decreasing temperature, showing considerable anisotropy between the two axes. At low temperature, there is a change in slope at around 12\,K along the two axes due to the reduction in spin disorder scattering of the charge carriers induced by the ordering of Eu$^{2+}$ magnetic moments. The drop in resistivity near 4\,K can be atributed to small traces of Sn flux, sticking superficially to the sample. Whether Sn flux is also responsible partly, for the appreciable anisotropy in the resistivity can not be ascertained from our data.

\section{Conclusion}

We have successfully grown single crystals of EuPtGe$_3$ by high temperature solution growth method using Sn as flux. The magnetic susceptibility, M\"ossbauer and specific heat data show the presence of divalent Eu ions, which undergo a transition to an antiferromagnetic structure below 11\,K, the transition presenting a small degree of first order. Contrary to EuPtSi$_3$, no cascade of transitions is observed in EuPtGe$_3$, which possesses an equal moment magnetic structure. This could be caused by the much smaller anisotropy in EuPtGe$_3$, which tends to favour equal moment structures. A mean field model with two antiferromagnetic sublattices is shown to reproduce some aspects of the magnetization data in the AF phase, and to derive the first neighbour exchange integral ${\cal J}_1 \simeq -1.72$\,K and the second neighbour integral ${\cal J}_2 = 0.67$\,K.

\end{document}